# Modulation of heterogeneous surface charge and flow pattern in electrically gated converging-diverging nanochannel

Movaffaq Kateb[a,b,*], Mohammadreza Kolahdouz[c], Morteza Fathipour[b,**]

[a] *BIOS lab on chip, MESA + Institute for Nanotechnology, University of Twente, Enschede, Netherlands*
[b] *MEMS & NEMS Lab, School of Electrical and Computer Engineering, University of Tehran, Tehran, Iran*
[c] *Nanoelectronic Center of Excellence, School of Electrical and Computer Engineering, University of Tehran, Tehran, Iran*

A B S T R A C T

The present study aims at utilizing field effect phenomenon to induce heterogeneous surface charge and consequently changing the fluid flow in a solid state nanochannel with converging-diverging periodicity. It is shown that the proposed geometry causes non-uniform radial field adjacent to channel walls which is stronger around the diverging section and weaker next to converging part of the wall. The later generates heterogeneous surface charge at channel walls depending on the applied gate potential i.e. applying low gate potential enables effective modulation of surface charge with the same polarity of the intrinsic charge at channel walls, while moderate gate potential causes charge inversion in diverging sections of the channel and generates reverse flow and thus results in fluid flow circulation. The potential application of flow circulation for trapping and rejection of particles is also demonstrated.

## 1. Introduction

Electroosmosis describes how an external electric field can initiate motion in an electrolytic fluid in the vicinity of immobile charged surface, and how its corresponding flow rate depends on surface characteristics. Recently electroosmotic flow in micro and nanochannel with non-uniform wall potential has been shown promising in mixing application [1]. In practice, the non-uniform surface potential may arise from surface defects or adsorption of analyte to the walls. However, the focus of the current study is on the locally altered surface potential to achieve specific flow patterns and its applications.

Anderson and Idol [2] initially studied electroosmotic flow in microchannel of periodically varying $\zeta$ potential and found out reversion of electroosmotic flow in different regions generates circulation. They also determined mean fluid velocity accurately by putting average $\zeta$ potential into Helmholtz-Smoluchowski equation ($\bar{u} = -\varepsilon E \zeta/\eta$). Ajdari [3,4] showed symmetric altering $\zeta$ potential with zero average gives only a circulation without net motion in axial direction. The later effect is sometimes considered as drawback due to reduced flowrates [5] but also has found use in passive mixers [1]. However, the tradeoff between mixing and transport is not resolved i.e. excellent mixing may lead to poor transport efficiency and vice versa [6].

The above-mentioned studies were assuming the thin EDL condition i.e. the ion concentration is uniform and thus the Poisson equation reduces to Laplace's equation. Ren and Li [7] using non-uniform $\zeta$ potential with overlapped EDL produced different types of the velocity profile in a microchannel. They also considered different $\zeta$ potential profiles along the channel and concluded that the flow rate is only determined by an average value of $\zeta$ potential independent of its distribution. The latter seems surprising since Helmholtz-Smoluchowski equation derived for thin EDL assumption. Thus, Fu et al. [8] highlighted the assumption of the Boltzmann distribution of charge density in these studies [1–7]. The Boltzmann model is only applicable at equilibrium i.e. constant electrochemical potential in the entire channel without any imposed electric field. Fu et al. [8] studied a step change in $\zeta$ potential by Nernst-Planck equation compared with classical Poisson–Boltzmann model. The former gives a gradual change of EDL around the $\zeta$ potential step while latter results in a sharp change in concentration which cannot be true due to convective transport of ions. They assumed zero $\zeta$ potential on one side of the channel which means no EDL, hence their solution could not predict circulation.

A more sophisticated control over $\zeta$ potential and fluid flow is offered by field effect i.e. applying a potential to a gate ($V_G$) which is separated from the fluid by a dielectric channel wall and affects $\zeta$ potential through capacitance according to Debye-Hückel theory. The later called ionic/fluidic field effect transistor (FET) in analogy to semiconductor MOSFET devices. Qian and Bau [9] proposed an analytical solution to FET structure consist of alternatively charged electrodes. Various flow patterns were obtained by time-wise variation of $\zeta$ potential e.g. overcome above mentioned mixing and transport tradeoff by sequential mix and transport. However, they only assumed a sharp variation of $\zeta$ potential instead of solving Poisson equation for the field effect. However, in practice, surface charge and $\zeta$ potential changes gradually between alternatively charged electrodes.

---

[*] Correspondence to: M. Kateb, Science Institute, University of Iceland, Dunhaga 3, IS-107 Reykjavik, Iceland.
[**] Corresponding author.
 *E-mail addresses:* mkk4@hi.is (M. Kateb), mfathi@ut.ac.ir (M. Fathipour).

On the other hand, experimental studies are restricted to available fabrication and characterization methods which are more evident at nanoscale. For instance, these studies are still limited to straight in-plane channels, with respect to the substrate, that eventually leads to low throughput devices [10,11].

In the present study, BOSCH process was used for the preparation of vertical (out-of-plane) channels in the large area. The sequential nature of BOSCH process leaves scalloped sidewalls which is usually considered as a drawback but in the present study the advantage of such a channel for generating heterogeneous $\zeta$ potential is exploited. The behavior of the device under different condition studied by simulation since out-of-plane channels do not allow imaging along the channel.

## 2. Method

### 2.1. Fabrication

The fabrication process began with deposition of a 70 nm thick low stress silicon nitride (SiRN) on a (100) p-silicon wafer. Then several 2 × 2 mm squares patterned in the wafaer's backside using photo-lithography followed by immersion in a hot phosphoric acid bath to open window in SiRN. Then the silicon was removed by KOH anisotropic wet etching through the window to make suspended SiRN/Si which becomes slightly transparent when the Si thickness reaches ~1 μm. The front face SiRN patterned by laser interference lithography [12] which followed by anisotropic etching in $SF_6$ and $C_4F_8$ mixture using inductively coupled plasma system to serve as hard mask. The BOSCH process consisted of 1, 1 and 2 s sequence of $C_4F_8$, $O_2$ and $SF_6$ at −40 °C and 1 kW was utilized for drilling silicon through the SiRN mask. The backside again was dry etched with $SF_6$, $O_2$ and $C_4F_8$ mixed at 1 kW to open pore bottoms and reach desired membrane thickness. Then an oxide was grown everywhere by dry oxidation at 945 °C to desired thickness.

### 2.2. The model

Fig. 1A shows the scanning electron microscope image of a fabricated channel consisting of periodic converging-diverging sections. Fig. 1B illustrates the geometry used in the simulation with channel walls and gate considered to be made of $SiO_2$ and Si, respectively. However, Si\-$SiO_2$ interface obtained in the fabrication process separated by $SiO_x$ layer i.e. interface charges with specific thickness which was neglected here. The average diameter ($D_{ave}$) of the channel assumed to be 130 nm which smoothly changes between 100 and 160 nm in the converging ($D_{min}$) and diverging ($D_{max}$) parts with a repetition period of 100 nm. It can be seen the $D_{min}$ was sharper in the fabricated channel. However, the fabricated channel with 200, 300 and 400 nm present more smooth edge at $D_{min}$ and thus sharp edges were avoided in the model for both gate and oxide. The gate and oxide thickness is minimized at $D_{max}$ and maximized at $D_{min}$ in the model.

### 2.3. Numerical approach

The EDL was modeled by solving mass transport of ions considering diffusion, ion-ion interaction and fluid convection as well as electromigration due to applied fields. These assumptions are of the prime importance since the contribution of converging-diverging geometry and field effect is still unknown and may alter charge distribution and EDL. The time dependent form of the Nernst-Planck equation is used to describe the motion of ionic species inside the channel.

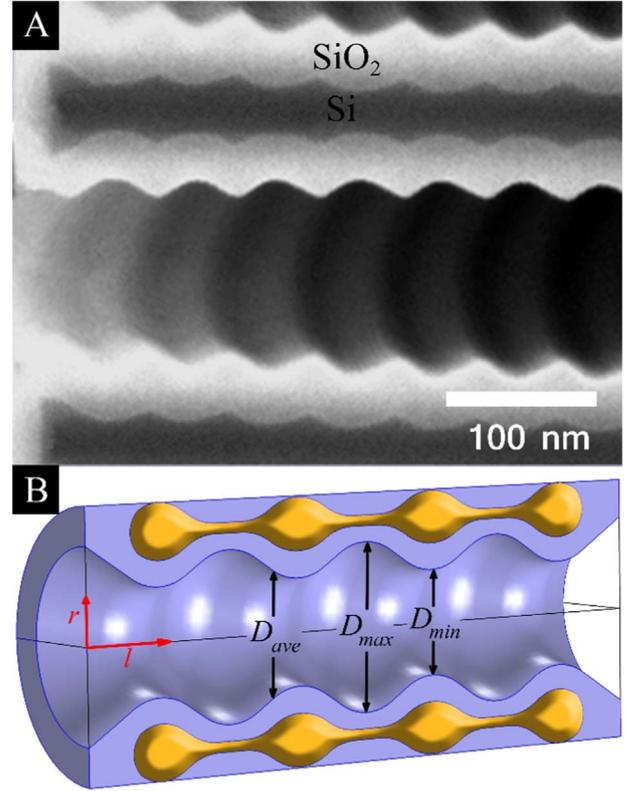

**Fig. 1.** (A) SEM image of the channel cross section obtained in the fabrication and (B) the model used in the simulation indicating $D_{min}$, $D_{ave}$ and $D_{max}$ as well as longitudinal, $l$, and radial, $r$, axis.

$$\frac{\partial c_i}{\partial t} + \boldsymbol{U}\cdot \nabla c_i = \nabla \cdot \left[ D_i \nabla c_i + \frac{D_i z_i e}{k_B T} c_i \nabla \phi \right] \quad (1)$$

where $c_i$, $D_i$ and $z_i$ are concentration, diffusion coefficient and ionic valence of the $i$ species. $\boldsymbol{U}$ is the velocity vector of the fluid and $\phi$ is the electric potential. The $e$, $k_B$ and $T$ respectively stand for elementary charge, Boltzmann constant and absolute temperature.

The mean field approximation of the electrostatic potential is described by the Poisson equation, which relates the electrical potential to the charge density.

$$\nabla^2 \phi = -\frac{\rho_{ch}}{\varepsilon} = -\frac{e}{\varepsilon} \sum_{i=1}^{n} z_i c_i \quad (2)$$

here $\rho_{ch}$ and $\varepsilon$ are the charge density and electric permittivity with $n$ being the total number of species in the system. We assumed similar $\varepsilon$ for the bulk electrolyte and EDL and neglected its variation with $V_G$ which is a valid assumption in non-overlapping EDL regime.

The Navier-Stokes equation is an expression of conservation of linear momentum for an incompressible Newtonian fluid with constant mass density. Now we allow for fluid motion due to electrostatic force by adding a term to the Navier-Stokes to represent the body force density due to the electrostatic force.

$$\rho_f \left( \frac{\partial \boldsymbol{U}}{\partial t} + \boldsymbol{U}\cdot \nabla \boldsymbol{U} \right) = -\nabla p + \eta \nabla^2 \boldsymbol{U} + \rho_{ch} \boldsymbol{E} \quad (3)$$

here $\boldsymbol{U}$, $\rho_f$, $\eta$ and $p$ respectively are the velocity vector, density, viscosity and pressure of the fluid. $\boldsymbol{E}$ is the electric field due to charge redistribution around the walls and the electric field applied along the channel as well as radial field imposed by $V_G$. An electric field of $10^5$ V/m was generated along the channel by applying proper voltage to inlet and outlet to generate electroosmotic flow.

Although no pressure gradient is imposed along the channel, neglecting the pressure, $p$, term means skipping osmotic pressure effect which is proportional to the charge density. Especially in the present configuration, the $\rho_{ch}$ is not constant throughout the channel and thus it changes $p$ abruptly next to channel walls. This is an important issue since it may lead to an artificial flow due to an unbalanced electrohydrostatic ion pressure stemming from the Maxwell stress term.

The left-hand side of the Eq. (3) represents the convective transfer of linear momentum which can be simplified by neglecting unsteady terms. This is an appropriate assumption unless there is forcing at high frequency. In addition, the continuity equation for an incompressible fluid leads to:

$$\nabla \boldsymbol{U} = 0 \tag{4}$$

The later assumption has been made for mixers of heterogeneous $\zeta$ potential with [9] and without [13] field effect (gate). For the case of non-uniform $\zeta$ potential, it has been demonstrated that the results of keeping [14] or skipping [8] the inertial term are in good agreement especially for ion distribution which determines EDL and electroosmotic flow.

The fluid considered to be water which is proper solvent for most of the electrolytes with a relative permittivity of 87.5 at room temperature. On each side of the channel, there is a reservoir with the constant concentration of electrolyte equal to 0.01 M KCl.

The particles were assumed to be rigid spheres of 10 (P$_1$) and 20 nm (P$_2$) in diameters, with the relative permittivity of 50 and 59. The density of both particles was chosen to be 1050 kg/m$^3$ and their electrical conductivity was assumed to be 1 μS/m with opposite charge. The EDL formed adjacent to the charged surface of a particle with a typical thickness ranging from 0.1 to 10 nm, is so thin that will not be resolved in detail. In the framework of the thin EDL approximation, the particles and their adjacent EDL were considered as a single entity, with 1 nm shell thickness and shell relative permittivity of 6 and 4.44 respectively for P$_1$ and P$_2$. This is a true assumption since the particles are small enough regarding the $D_{min}$ of the channel. For bigger particles, however, EDL around particles cannot be neglected [15,16].

The coupled system described above is simultaneously solved with a commercial finite-element package COMSOL (version 5.2) and Matlab R2016a. In order to validate the present computational method, we made comparisons with existing analytical and experimental results of electroosmotic flow in channels with a simple flat geometry.

## 3. Results and discussion

### 3.1. Device characteristics

Fig. 2 illustrates the regulation of net charge density, $\rho_{ch}$, radial electric field, $E_r$, and potential, $\phi$, at two different part of the channel along $D_{max}$ and $D_{min}$. The figure also contains the result of a cylinderical channel ($D_{cylinder}$) with 100 nm in diameter for comparison. The fluid, oxide and gate regeions are also labeled and separated by a yellow highlight in the figure for more clarity. It can be seen that the channel presents characteristic properties of MOSFET devices. For instance, zero $\rho_{ch}$ in the bulk fluid which changes in the EDL followed by a sharp change to zero in the oxide. However, the thickness variation of the EDL seems to be negligible and thus we are in non-overlapping EDL regime. In analogy to a semiconductor MOSFET, which the variation of depletion zone width reduces at higher dopant concentration, here the high strength of electrolyte reduces EDL thickness variation. It can be seen that at zero $V_G$, the same charge density was obtained at the wall in $D_{max}$, $D_{min}$ and $D_{cylinder}$ which will be called *intrinsic surface charge* hereafter. The effect of complex morphology is more pronounced at $V_G$ of 0.6 V which present inverse polarity of EDL at $D_{max}$ compared to $D_{min}$ and $D_{cylinder}$. Further increase in $V_G$, would results in similar polarity at $D_{max}$, $D_{min}$ and $D_{cylinder}$. It is also worth noting that the $D_{cylinder}$ presents an intermediate charge between values at $D_{max}$ and $D_{min}$ which indicates the importance of curvature.

Since there is no charge and charge transfer within the oxide, the $E_r$ is expected to be constant which is the case for the cylinder. However, in the current geometry the field lines starting from the gate are converging toward $D_{max}$ and thus the magnitude of $E_r$ increases gradually up to the channel walls. Inversely, the filed lines become diverging toward $D_{min}$ that cusses reduction in the magnitude of $E_r$.

The $\phi$ equals to $V_G$ in the gate which drops drastically across the oxide but a slight change in $\phi$ at the channel wall is enough to change the electroosmotic flow in the channel. Since $E_r$ is not constant, the $\phi$ drop within the oxide shows deviation from linear behavior obtained in the cylinder.

### 3.2. Space charge density

Fig. 3 illustrates the distribution of $\rho_{ch}$ at different applied $V_G$ in the side view cross section of the channel. At zero $V_G$, accumulation of cations occurs at the channel walls due to intrinsic surface charge ($\sigma_0$). For $V_G$ of 0–0.4 V, not shown here, the walls are still surrounded by cations but the $\rho_{ch}$ values drops slightly around the channel walls. Applying 0.5 V to the gate overcomes $\sigma_0$ at channel walls but only around $D_{max}$ which causes anion accumulation as indicated by deep blue. While cations still pile up around $D_{min}$ which is evident from the red line and cyan halation. The $V_G$ of 0.6 V almost balances the density of anions and cations at $D_{max}$ and $D_{min}$, respectively. This effect enables efficient mixing by inversion of electroosmotic flow in converging and diverging part of the channel. Slightly higher $V_G$ (0.7 V), unbalances the $\rho_{ch}$ and increases anions density around $D_{max}$ and this trend continues by further increase in $V_G$. It is worth noting that the application of $V_G$ cannot provide uniform anions density along the channel walls since the strength of the field is different at $D_{max}$ and $D_{min}$.

To illustrate net charge distribution more quantitively, the exact value of $\rho_{ch}$ for various $V_G$ is plotted in Fig. 4. At $V_G$ equazero, the $\rho_{ch}$ is positive in all cases as result of $\sigma_0$ those shift toward negative values by applying positive $V_G$. At $V_G \sim 0.6$ V the $\rho_{ch}$ presents reversed values at $D_{max}$ and $D_{min}$.

The point of zero charge (PZC) is obtained at different gate voltages namley around 0.4, 0.6 and 0.9 V respectively for $D_{max}$, $D_{ave}$ and $D_{min}$. In other words, for $V_G$ between 0.4 and 0.9 V a specific part of the channel between $D_{max}$ and $D_{min}$ is at PZC. The latter is practically important since the maximum solubility of the gate oxide occurs at PZC.

The ionic inversion voltage, compared to the $V_G = 0$, is around 0.8, 1.2 and 1.8 V at $D_{max}$, $D_{ave}$ and $D_{min}$, respectively. The possibility of ionic inversion obtained here called ambipolar effect. This is in contradiction to the results of Karnik et al. [10] which claimed ambipolar effect is not accessible using low dielectric constant oxide such as SiO$_2$. Thus, they asserted that the efficiency of FET device is limited to tuning the charge with the same polarity of intrinsic charge e.g. cation density at the channel walls. However, they used a planar gate on one side of the channel which means one side of the channel experienced applied field efficiently. Later studies by Lee et al. [11] showed a surrounded gate structure enables charge conversion at channel walls or so called ambipolar effect. Then the ambipolarity of fluidic FET is highly dependent on the gate structure and how efficiently it applies the field into the channel.

### 3.3. Zeta potential

Variation of the $\zeta$ potential with the $V_G$ is shown in Fig. 5 which depicts the characteristic non-linear behavior of $\zeta$ potential [17]. However, the deviation is bigger for $D_{max}$ and it saturates at higher $\zeta$ potential which can be explained by the difference in capacitance as follows:

Assuming the EDL and the gate oxide as parallel plates the Gauss' law gives:

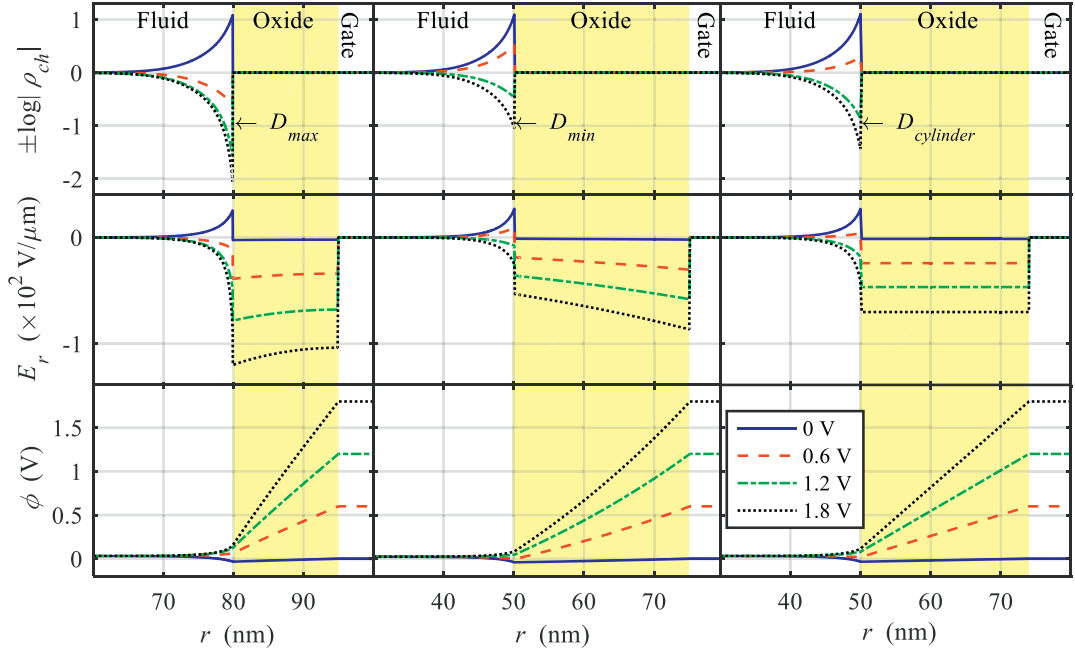

**Fig. 2.** Radial variation of charge density, electric field and potential at two different part of channel along $D_{max}$ and $D_{min}$ in comparison with a flat cylinder ($D_{cylinder}$).

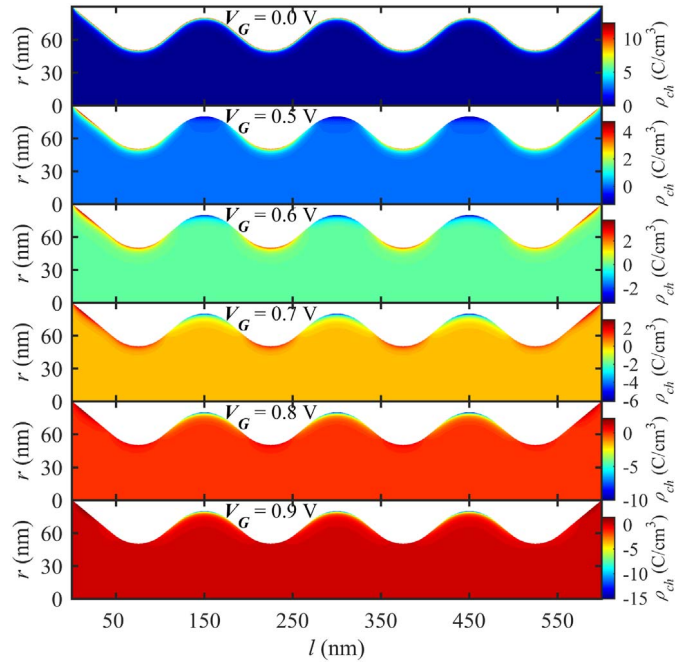

**Fig. 3.** Charge distribution at different $V_G$ of 0–0.9 V in the side view cross section of the channel.

$$\varepsilon_f E_r^f - \varepsilon_{ox} E_r^{ox} = \frac{\sigma}{\varepsilon_0} \quad (5)$$

where $ox$ and $f$ superscript denotes $E_r$ in the gate oxide and fluid respectively. $\varepsilon_0$ is the permittivity of the vacuum, $\varepsilon_{ox}$ and $\varepsilon_f$ are relative permittivity of the gate oxide and fluid and $\sigma$ is the surface charge density on the channel-liquid interface.

Assuming that the potential in the bulk is zero, Eq. (5) can be rewritten as,

$$C_d \phi - C_{ox}(V_G - \phi) = \sigma \quad (6)$$

where $C_{ox} = \varepsilon_0 \varepsilon_{ox}/t_{ox}$ is the capacitance per unit area of the oxide layer having thickness $t_{ox}$, and $C_d = d\sigma/d\phi$ is the differential capacitance. As illustrated by Grahame [18], the Gouy-Chapman- Stern model divides $C_d$ to the capacitance of the charges held at the outer Helmholtz layer ($C_H$) and capacitance of the truly diffuse charge ($C_{diff}$). The former is independent of $\phi$, but the later varies in a v-shaped fashion with the $\phi$. The composite capacitance shows a complex behavior and is governed by the smaller of the two components. At larger electrolyte concentrations, or even at large polarizations in dilute media, $C_{diff}$ becomes so large that it no longer contributes to $C_d$ and one sees only the constant capacitance of $C_H$. However, in general, $C_{diff}$ is smaller and shows the major contribution to the total capacitance in series. Here, the EDL capacitance assumed to be equal to that of the diffused layer.

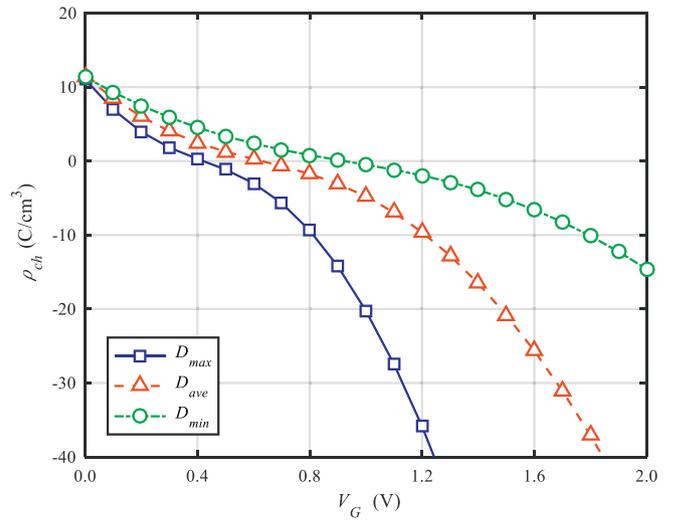

**Fig. 4.** Variation of $\rho_{ch}$ versus $V_G$ at the $D_{max}$, $D_{ave}$ and $D_{min}$ adjacent to channel walls.

$$\zeta = \frac{\sigma + C_{ox} V_G}{(C_{ox} + C_{diff})} \quad (7)$$

When there is no voltage on the gate:

$$\zeta_0 = \frac{\sigma}{(C_{ox} + C_{diff})} \quad (8)$$

where $\zeta_0$ is the intrinsic $\zeta$ potential of the channel wall. Now rewriting

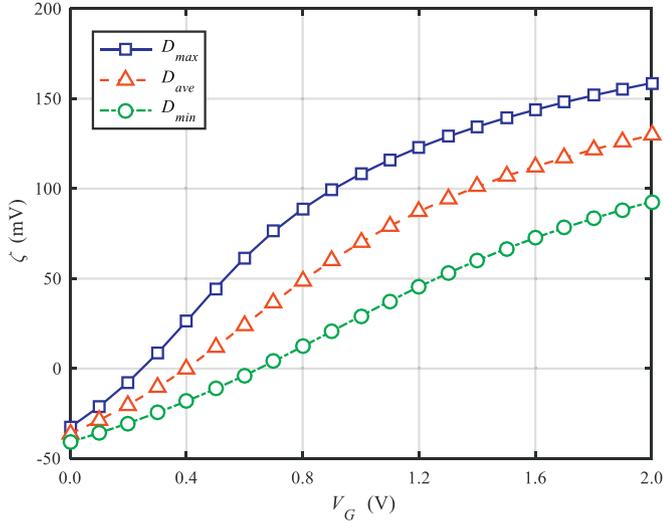

**Fig. 5.** Variation of $\zeta$ potential with $V_G$ at $D_{max}$, $D_{ave}$ and $D_{min}$.

Eq. (7) gives:

$$\zeta = \frac{C_{ox}}{(C_{ox} + C_{diff})} V_G + \zeta_0 \qquad (9)$$

Eq. (9) seems to predict liner variation for $\zeta$ potential but this relation is more complicated since $C_{diff}$ would change with $\zeta$ potential. However, this simplified model enables describing the difference in Fig. 5. The EDL around $D_{max}$ experience stronger applied field which results in dielectrophoretic saturation of counter ions and reduction of $\varepsilon_f$. Consequently, capacitance becomes smaller at $D_{max}$ and the $\zeta$ potential increases.

### 3.4. Velocity field

Fig. 6 shows the axial velocity, $U_l$, profile in the cross section of the channel at $D_{max}$ and $D_{min}$ for a range of $V_G$ (0–2 V). It can be seen that applying proper $V_G$ enables efficient control over the fluid flow at both $D_{max}$ and $D_{min}$. However, in both cases at $V_G$ of 0.8 V, and 0.6 V along $D_{max}$, the $U_l$ is reversed next to the wall with respect to center as result of ionic inversion. The later generates circulation in both cases and is of interest for efficient mixing.

Away from moderate $V_G$, the velocity profile shows a deviation from the ideal plug flow i.e. constant $U_l$ in the bulk with a drop to zero in EDL within a flat channel. Here the velocity profile reaches a maximum magnitude in the EDL with a drop toward the center of the channel. Such a profile has been reported for transient development of the plug flow in a flat micro-slit [19]. However, in the converging-diverging channel, the fluid always experiences variation in cross section and thus stable profile would never become flat in the bulk. It is worth noting that the velocity drop in the bulk is more pronounced along $D_{max}$. This is associated with the combination of transient plug flow and continuity since the latter compensates for the decrease in cross section.

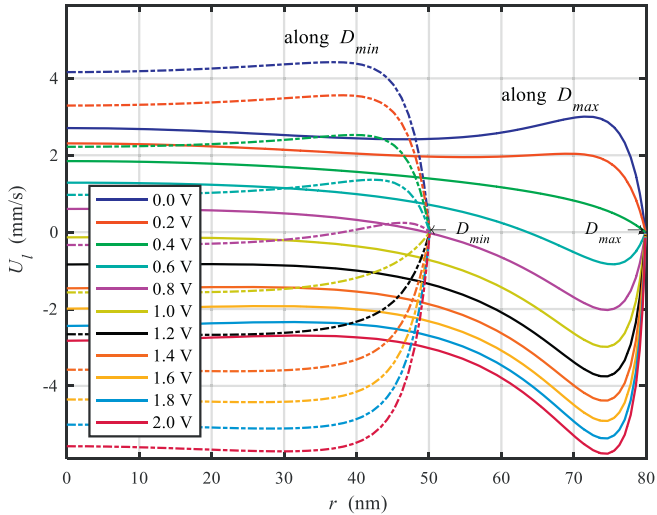

**Fig. 6.** Variation of axial velocity profile with $V_G$ along $D_{max}$ (line) and $D_{min}$ (dash-dot line).

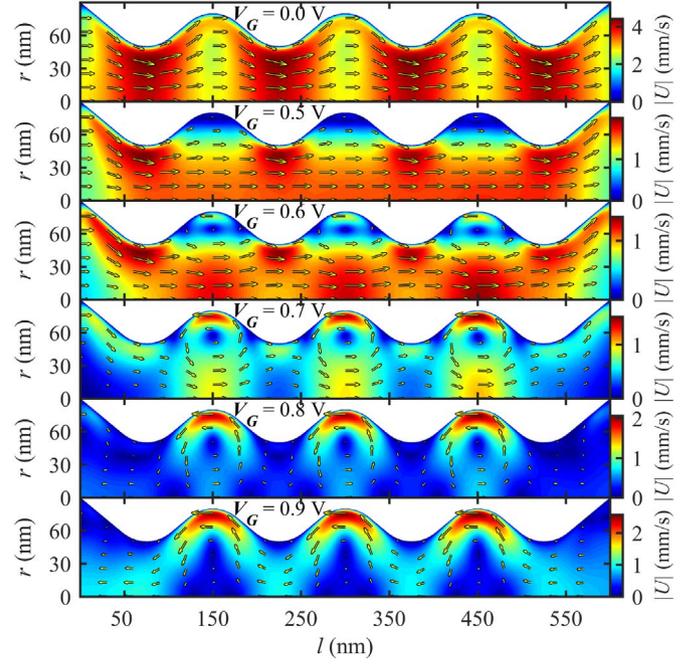

**Fig. 7.** The magnitude of velocity field in side view cross section of the channel at different $V_G$. The arrows and indicating the flow direction.

To provide more evidence of circulation, electroosmotic velocity in the side view cross section of the channel is shown in Fig. 7 at several $V_G$. In the lack of $V_G$, the maximum velocity occurs in converging sections due to continuity Eq. (4). Thus, a moving particle experiences acceleration and deceleration in the red and yellow regions, respectively. At $V_G$ of 0.5 V, as result of slight anion pile up around $D_{max}$ a slow current circulation occurs with zero velocity in the middle of the blue region. The red region is indicating the maximum velocity also confined close to $D_{min}$ where the cations are still available. It is evident that the velocity is uniform in the center of the channel ($r$ equal to 0–15 nm) and thus there would be no acceleration-deceleration behavior in the center. As discussed before, a strong enough field causes anion accumulation at $D_{max}$ channel walls which eventually generates considerable reverse flow. This is shown in $V_G$ of 0.6 V where reverse flow forms a vortex in the diverging section with zero velocity in the vortex center. The vortex act as an obstacle that decreases the channel cross section and results in the maximum velocity below the vortex in the center of the channel. This is very interesting since around the center of the channel, acceleration and deceleration is swapped compared to $V_G = 0$. However, it is doubted whether velocity field in the rest of the channel guides the particle into the vortex or not. $V_G$ of 0.7 V, generates the maximum velocity around the $D_{max}$ by providing a higher density of anion. It also causes further displacement of the vortex center toward the center of the channel. The higher $V_G$ of 0.8 V holds the maximum velocity around $D_{max}$ while moves the vortex center even more and develops a very small reverse flow to the left in the blue regions. Further increase in $V_G$, as shown in $V_G = 0.9$ V subplot, forms a serpent like reverse flow and extends the vortex center toward the center of the channel forming those blue triangles.

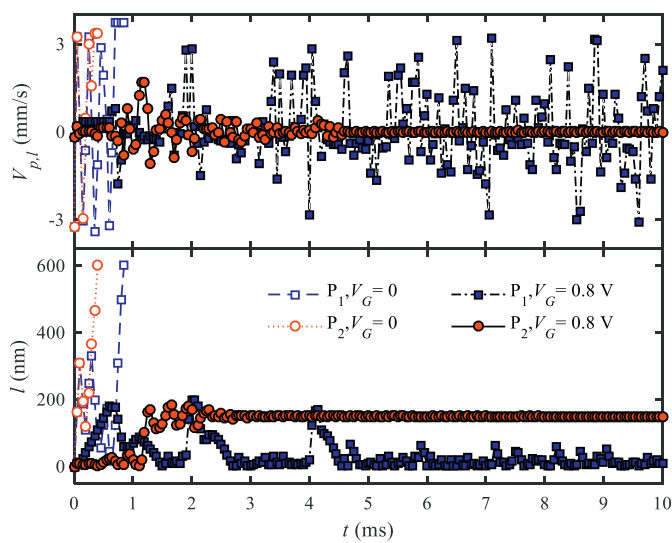

**Fig. 8.** Variation of particle velocity and position along the channel with elapsed time.

*3.5. Particle tracing*

Fig. 8 compares the velocity and position along the channel for two particles of opposite charge at different $V_G$. At zero $V_G$, depicted by hollow symbols, the velocity shows fluctuation between ± 3 mm/s and reaches a maximum till particles leave the channel i.e. when $l = 600$ nm in the lower plot. The time for reaching the maximum velocity is 0.7 and 0.35 ms for $P_1$ and $P_2$, respectively. At $V_G$ of 0.8 V shown by filled symbols, the $P_1$ is rejected which is evident from its position at channel inlet ($l = 0$) with its velocity become more unstable at the longer time. While for $P_2$ the velocity fluctuates at the very beginning and nearly converges to zero which means it is trapped at $l = 150$ according to the lower plot.

## 4. Conclusion

In conclusion, we have introduced a simple and large area technique for fabrication of ionic FET device on the basis of using nanoscale patterning and DRIE followed by dry oxidation. The converging-diverging geometry was obtained as result of the sequential nature of DRIE which remains unchanged after oxidation. The field effect control was achieved through the gate electrode which surrounded the outer surface of $SiO_2$ channel walls to generate radial electric field and effectively modulate the surface charge at the channel-liquid interface.

The results show the application of $V_G$ generates heterogeneous surface charge in diverging and converging sections. This field induced redistribution of ions affects the electrokinetic transport of fluids and suspended particles in the channel. The intermediate $V_G$ causes circulation of the fluid which can be utilized for efficient mixing on its own. This is very promising since previous mixers hold complicated structure, with at least two adjacent gate electrodes, while present structure eliminates multi-electrodes and suggests simple fabrication techniques. By introducing particles, the circulation causes different behaviors of focusing, trapping and rejection at different gate voltages.

## Acknowledgment

Authors would like to thank Henk van Wolferen and Johan Bomer respectively from nanolab and BIOS in mesa + institute who shared their expertise in nanofabrication. We would like to thank our colleagues Amirali Ebadi, Mohsen Hajari and Sedigheh Nikpay in the MEMS & NEMS lab that greatly assisted the initial simulations.